\documentclass[a4paper,twoside,12pt]{article}
\usepackage{amsmath,amsfonts,amsthm,graphicx}
\numberwithin{equation}{section} 
\newtheorem{theorem}{Theorem}[section]
\newtheorem{conjecture}[theorem]{Conjecture}
\begin{document}
\title{Integral equations for thermodynamics of the 
$osp(1|2s)$ integrable spin chain}
\author{ 
Zengo Tsuboi
\footnote{E-mail address: ztsuboi@poisson.ms.u-tokyo.ac.jp}
\\
{\it Graduate School of Mathematical Sciences, 
University of Tokyo,} \\ 
{\it Komaba 3-8-1, Meguro-ku, Tokyo 153-8914, Japan}
}
\date{}
\maketitle
\begin{abstract}
We propose a system of nonlinear integral equations (NLIE), 
which gives the free energy of the 
$osp(1|2s)$ integrable spin chain at finite temperatures. 
 In contrast with usual thermodynamic Bethe ansatz equations, 
 our new NLIE contain only a {\em finite} number of 
 unknown functions. 
 On deriving NLIE, we use our 
 $osp(1|2s)$ version of the $T$-system and 
the quantum transfer matrix method. 
Based on our NLIE, we also calculate 
the high temperature expansion of the 
free energy and the specific heat. 
\end{abstract}
{\it PACS2001:} 02.30.Rz; 02.30.Ik; 05.50.+q; 05.70.-a 
{\bf Physics Letters B 544 (2002) 222-230.} 
%
\section{Introduction}
In recent years,  
thermodynamics of solvable lattice models 
related to superalgebras 
have been studied by thermodynamic Bethe ansatz (TBA) method, 
and TBA equations are examined from various point of view 
(see, section 1 in \cite{T02} for a comment on the 
present status of TBA equations for models related 
to superalgebras). 
In our previous papers, we have derived TBA equations for the 
$osp(1|2)$ model \cite{ST99,ST00,ST01} and the $osp(1|2s)$ 
model \cite{T02} from the string hypothesis 
and the $T$-system \cite{T99} of the quantum transfer matrix 
(QTM). 
These TBA equations contain an infinite number of unknown functions, 
 and thus are not always easily treated.  
It is important to reduce the TBA equations to tractable 
integral equations which contain only a 
finite number of unknown functions. 

As for the $XXZ$ spin chain, which is related to the algebra 
$U_{q}(A^{(1)}_{1})$ of rank one, 
 Takahashi proposed \cite{Ta01} a nonlinear integral equation 
 (NLIE) recently. 
This NLIE contains only one unknown function. 
Due to its simplicity, we can calculate \cite{ShT02} 
the high temperature expansion of physical quantities 
to very high order from this NLIE. 
The purpose of this letter is to derive a system of NLIE 
with only a {\em finite} number (the number of rank $s$) 
of unknown functions 
from our $osp(1|2s)$ version of the $T$-system \cite{T99}. 
This letter is the first attempt to derive this type of NLIE 
for a vertex model associated with an algebra of {\em arbitrary} rank 
\footnote{In the case of the deformation parameter $q$ 
of an underlining quantum affine algebra 
is root of unity ($|q|=1,q \ne 1$), the $T$-system becomes a finite 
set of difference equations. Thus the corresponding TBA equation 
becomes a finite set of integral equations. 
See, \cite{SW02} for TBA analysis of integrable field theories 
related to $osp(1|2s)$. 
We also note that different types of 
NLIE with finite numbers of unknown functions 
for different algebras of arbitrary rank were considered 
in \cite{Z98,DDT00} in rather different contexts.
}. 

In section 2, we briefly mention the $T$-system 
for the $osp(1|2s)$ model and the QTM method 
\cite{S85,SI87,K87,SAW90,Kl92}. 
This section overlaps with section 2 and section 4 in \cite{T02}, but  
normalization of the fused QTM is different from \cite{T02}. 
In section 3, we derive the NLIE. 
The normalized QTM defined in section 2 play the role of 
the unknown functions of these NLIE. 
These NLIE contain a parameter $m$, which corresponds 
to the fusion degree of the model. 
For $m=1$, these NLIE form a closed set of equations, 
which contains only a finite number of unknown functions. 
On the other hand for $m \ge 2$, 
they couple with the ones for $m=1$, and contain 
 an infinite number of unknown functions. 
This type of equations (for $m \ge 2$) has never been considered before 
even in the case of Takahashi's 
NLIE for the $XXZ$ spin chain \cite{Ta01}. 
These NLIE relate to traditional TBA equations 
 through the dependant variable transformation (\ref{Y-fun}). 
Using our new NLIE, we calculate the 
high temperature expansion of the free energy in section 4. 
Section 5 are devoted to discussions.
\section{$T$-system and QTM method}
In this section, we consider \cite{T02} an integrable 
spin chain\cite{Mar95-2,MR97} associated with the 
fundamental representation of $osp(1|2s)$, 
and introduce 
the QTM \cite{S85,SI87,K87,SAW90,Kl92} 
and the $T$-system \cite{T99} for this model. 
The $\check{R}$-matrix\cite{Kul86,BS88,ZBG91,MR94,MR97} 
of this model is given as 
\begin{eqnarray}
\check{R}(v)=I+vP-\frac{2v}{2v-g}E, \label{R-mat}
\end{eqnarray}
where $g=2s+1$; $v\in {\mathbb C}$; 
$P^{cd}_{ab}=(-1)^{p(a)p(b)}\delta_{ad}\delta_{bc}$; 
$E^{cd}_{ab}=\alpha_{ab}(\alpha^{-1})_{cd}$; 
$a,b,c,d 
\in \{1,2,\dots,s,0,\overline{s},\dots , \overline{1}\} $; 
$ 1 \prec 2 \prec \cdots \prec s \prec 0 \prec 
\overline{s} \prec \cdots \prec \overline{2} \prec \overline{1}$; 
$\alpha$ is $(2s+1)\times (2s+1)$ anti-diagonal matrix whose 
non-zero elements are  
$\alpha_{a,\overline{a}}=1$ for $a \in \{1,2,\dots,s,0\}$ and  
$\alpha_{a,\overline{a}}=-1$ for 
$a \in \{ \overline{s},\overline{s-1},\dots , \overline{1} \} $; 
$\overline{\overline{a}}=a$; 
$p(a)=0$ for $a=0$; $p(a)=1$ for 
$a \in \{1,2,\dots,s\} \sqcup  
\{\overline{s},\dots,\overline{2},\overline{1}\}$. 
The Hamiltonian of the present 
model for the periodic boundary condition is given by 
\begin{eqnarray}
H=J\sum_{k=1}^{L}\left(P_{k,k+1}+\frac{2}{g}E_{k,k+1}\right),
\label{Hamiltonian}
\end{eqnarray}
where $J$ is a coupling constant: $J>0$ and $J<0$ correspond to 
the ferromagnetic and antiferromagnetic regimes respectively; 
$L$ is the number of the lattice sites; $P_{k,k+1}$ and 
$E_{k,k+1}$ act nontrivially 
on the $k$ th site and $k+1$ th site. 

The QTM is defined as 
\begin{eqnarray}
t^{(1)}_{1}(v)={\mathrm Tr}_{j}\prod_{k=1}^{\frac{N}{2}}
 R_{a_{2k},j}(u+iv)\widetilde{R}_{a_{2k-1},j}(u-iv),
 \label{QTM}
\end{eqnarray}
where $R^{cd}_{ab}(v)=\check{R}^{cd}_{ba}(v)$; 
 $\widetilde{R}_{jk}(v)=^{t_{k}}\! \! R_{kj}(v)$ 
($t_{k}$ is the transposition in 
the $k$-th space); $N$ is the Trotter number and assumed to be even. 
By using the largest eigenvalue $T^{(1)}_{1}(0)$ of 
the QTM (\ref{QTM}) at $v=0$, 
the free energy per site is expressed as 
\begin{eqnarray}
f=
-T\lim_{N\to \infty}\log T^{(1)}_{1}(0),
\end{eqnarray}
where we set $u=-\frac{J}{T N}$ ($T$: temperature); 
the Boltzmann constant is set to $1$. 
One can obtain the eigenvalue formulae of the QTM (\ref{QTM}) by 
replacing the vacuum part of the dressed vacuum form 
(DVF) for the row-to-row transfer matrix with that of the QTM. 
This DVF is imbedded into a DVF for a fusion hierarchy of the QTM. 
It reads as follows. 
\begin{eqnarray}
T_{m}^{(a)}(v)=\sum_{\{d_{jk}\}} \prod_{j=1}^{m}\prod_{k=1}^{a}
z(d_{jk};v-\frac{i}{2}(m-a-2j+2k)),
\label{DVF}
\end{eqnarray}
where the summation is taken over 
$d_{jk}\in 
\{1,2,\dots,s,0,\overline{s}, \dots, \overline{2},\overline{1}\}$ 
such that $d_{jk} \preceq d_{j+1,k}$ and $d_{jk} \prec d_{j,k+1}$. 
The functions $\{z(a;v)\}$ are defined as
\begin{eqnarray}
&& z(a;v)=\psi_{a}(v)
\frac{Q_{a-1}(v+\frac{i}{2}(a+1))Q_{a}(v+\frac{i}{2}(a-2))}
{Q_{a-1}(v+\frac{i}{2}(a-1))Q_{a}(v+\frac{i}{2}a)} 
\nonumber \\
&& \hspace{150pt} \mbox{for} \quad a \in \{1,2,\dots,s\}, 
\nonumber \\
&& z(0;v)=\psi_{0}(v)
\frac{Q_{s}(v+\frac{i}{2}(s-1))Q_{s}(v+\frac{i}{2}(s+2))}
     {Q_{s}(v+\frac{i}{2}(s+1))Q_{s}(v+\frac{i}{2}s)}, \\  
&& z(\overline{a};v)=\psi_{\overline{a}}(v)
\frac{Q_{a-1}(v-\frac{i}{2}(a-2s))Q_{a}(v-\frac{i}{2}(a-2s-3))}
{Q_{a-1}(v-\frac{i}{2}(a-2s-2))Q_{a}(v-\frac{i}{2}(a-2s-1))} 
\nonumber \\
&& \hspace{150pt} \mbox{for} \quad a \in \{1,2,\dots,s\},
\nonumber 
\end{eqnarray}
where $Q_{a}(v)=\prod_{k=1}^{M_{a}}(v-v_{k}^{(a)})$; 
$M_{a}\in {\mathbb Z}_{\ge 0}$; $Q_{0}(v):=1$.  
The vacuum parts are given as follows
\begin{eqnarray}
\psi_{a}(v)=
\left\{
 \begin{array}{lll}
 \zeta_{1}\frac{\phi_{+}(v)\phi_{-}(v+i)\phi_{+}(v-\frac{2s-1}{2}i)}
   {\phi_{+}(v-\frac{2s+1}{2}i)}  &
 \mbox{for} & a=1,\\ 
 \zeta_{a}\phi_{+}(v)\phi_{-}(v)
 &
 \mbox{for} & 2 \preceq a \preceq \overline{2}, \\ 
 \zeta_{\overline{1}}\frac{\phi_{-}(v)\phi_{+}(v-i)\phi_{-}(v+\frac{2s-1}{2}i)}
   {\phi_{-}(v+\frac{2s+1}{2}i)}
 & \mbox{for} & a=\overline{1}, 
 \end{array}
\right.
\label{vac-QTM}
\end{eqnarray}
where $\phi_{\pm}(v)=(v\pm iu)^{\frac{N}{2}}$; 
 $\zeta_{a}$ is a phase factor. 
$\{v^{(a)}_{k}\}$ is a solution of the Bethe ansatz equation 
(BAE)
\begin{eqnarray}
\left\{
\frac{\phi_{-}(v^{(a)}_{k}+\frac{i}{2})
      \phi_{+}(v^{(a)}_{k}+\frac{i}{2}-\frac{ig}{2})}
     {\phi_{-}(v^{(a)}_{k}-\frac{i}{2})
      \phi_{+}(v^{(a)}_{k}-\frac{i}{2}-\frac{ig}{2})}
\right\}^{\delta_{a 1}}=
-\varepsilon_{a}
\prod_{d=1}^{s+1}
\frac{Q_{d}(v^{(a)}_{k}+\frac{i}{2}B_{ad})}
     {Q_{d}(v^{(a)}_{k}-\frac{i}{2}B_{ad})},
     \label{BAE}
\end{eqnarray}
where $k\in \{1,2, \dots, M_{a}\}$; $a\in \{1,2, \dots, s\}$; 
$Q_{s+1}(v):=Q_{s}(v)$;  
$B_{ad}=2\delta_{ad}-\delta_{a,d+1}-\delta_{a,d-1}$; 
  $\varepsilon_{a}$ is a phase factor. 
For $a \in \{1,2,\dots,s\}$ and $m \in {\mathbb Z}_{\ge 1}$, we 
 define a normalization function
\begin{eqnarray}
\hspace{-30pt} && \widetilde{{\mathcal N}}^{(a)}_{m}(v)=
  \frac{\phi_{-}(v+\frac{m+a}{2}i)\phi_{+}(v-\frac{m+a}{2}i)}{
  \phi_{-}(v-\frac{m-a}{2}i)\phi_{+}(v+\frac{m-a}{2}i)}
  \nonumber \\ 
\hspace{-30pt}  && \hspace{20pt} \times
  \prod_{j=1}^{m}\prod_{k=1}^{a}
  \phi_{-}(v-\frac{m-a-2j+2k}{2}i)\phi_{+}(v-\frac{m-a-2j+2k}{2}i), 
  \label{normal}
\end{eqnarray}
 and set 
 $ \widetilde{T}^{(a)}_{m}(v)=
 T^{(a)}_{m}(v)/\widetilde{{\mathcal N}}^{(a)}_{m}(v)$. 
For $s=1$, $\widetilde{T}^{(1)}_{1}(v)$ coincides with 
 eq.(4.8) in  \cite{ST00}. 
The poles of $\widetilde{T}^{(a)}_{m}(v)$ 
from the functions $\{Q_{b}(v) \}$ (dress part) 
are spurious under the BAE (\ref{BAE}).  
If one formally set the vacuum parts (\ref{vac-QTM}) and (\ref{normal}) 
to 1, then the remaining part 
(dress part) of the DVF $\widetilde{T}^{(a)}_{m}(v)$ is same as  
the row-to-row case. Thus $\widetilde{T}^{(a)}_{m}(v)$ satisfies 
the following functional relation, 
which has essentially the same form as the $osp(1|2s)$ 
$T$-system in \cite{T99}. 
\begin{eqnarray}
&& \hspace{-20pt}
\widetilde{T}^{(a)}_{m}(v+\frac{i}{2})
\widetilde{T}^{(a)}_{m}(v-\frac{i}{2})
=\widetilde{T}^{(a)}_{m+1}(v)\widetilde{T}^{(a)}_{m-1}(v)
+\widetilde{T}^{(a-1)}_{m}(v)\widetilde{T}^{(a+1)}_{m}(v),
\nonumber \\ 
&& \hspace{-20pt}
\widetilde{T}^{(s)}_{m}(v+\frac{i}{2})
\widetilde{T}^{(s)}_{m}(v-\frac{i}{2})
=\widetilde{T}^{(s)}_{m+1}(v)\widetilde{T}^{(s)}_{m-1}(v)+
\widetilde{T}^{(s-1)}_{m}(v)\widetilde{T}^{(s)}_{m}(v) \nonumber \\
&& \hspace{70pt} 
{\rm for} \quad a \in \{1,2,\dots,s-1\}
\quad {\rm and} \quad m \in {\mathbb Z}_{\ge 1}, 
\label{T-system}
\end{eqnarray}
where 
\begin{eqnarray}
\widetilde{T}^{(a)}_{0}(v)&=&1
\quad {\rm for} \quad a \in {\mathbb Z}_{\ge 1},\nonumber \\
\widetilde{T}^{(0)}_{m}(v)&=&
 \frac{\phi_{-}(v-\frac{m}{2}i)\phi_{+}(v+\frac{m}{2}i)}
  {\phi_{-}(v+\frac{m}{2}i)\phi_{+}(v-\frac{m}{2}i)}
\quad {\rm for} \quad m \in {\mathbb Z}_{\ge 1}. 
\end{eqnarray}
\section{Nonlinear integral equations}
In  \cite{TSK01}, Takahashi's NLIE 
for the $XXZ$-model \cite{Ta01} was derived from 
the $T$-system of the QTM. 
Here we derive our new NLIE from 
our $T$-system (\ref{T-system}). 

A numerical analysis for finite $N,u,s$ indicates that 
a two-string solution (for every color) in the sector 
$N=M_{1}=M_{2}=\cdots =M_{s}$ of the BAE (\ref{BAE}) provides 
the largest eigenvalue of the QTM (\ref{QTM}) at $v=0$ \cite{T02}. 
From now on, we consider only this two-string solution. 
In this case, the phase factors are 
$\varepsilon_{a}=\zeta_{a}=1$ for any $a$. 
Moreover, we expect the following conjecture is valid for 
this two-string solution \cite{T02}. 
\begin{conjecture}\label{conj}
For small $u$ ($|u|\ll 1$), $a \in \{1,2,\dots,s\}$ and 
$m \in {\mathbb Z}_{\ge 1}$,
every zero $\{\tilde{z}^{(a)}_{m} \}$ 
of $\widetilde{T}^{(a)}_{m}(v)$ is located near the lines 
${\mathrm Im} v= \pm \frac{m+a}{2}, \pm \frac{g+m-a}{2}$.
\end{conjecture}
$\widetilde{T}^{(a)}_{m}(v)$ has poles only at 
$\pm \tilde{\beta}^{(a)}_{m,k}$ ($k=1,2$): 
$\tilde{\beta}^{(a)}_{m,1}=\frac{m+a}{2}i -iu$, 
$\tilde{\beta}^{(a)}_{m,2}=\frac{g+m-a}{2}i -iu$. 
These poles are of order $N/2$ at most. 
Moreover  
$\lim_{|v|\to \infty}\widetilde{T}^{(a)}_{m}(v)=Q^{(a)}_{m}$ 
is a finite number 
\begin{eqnarray}
Q^{(a)}_{m}=\left(\frac{(m+g)!m!}{(m+a)!(m+g-a)!}\right)^m
 \prod_{k=1}^{m}\left\{ \frac{(k+a)(k+g-a)}{k(k+g)} \right\}^{k}.
 \label{q-sys-sol}
\end{eqnarray}
This is a solution of the $Q$-system \cite{T02}: 
\begin{eqnarray}
&&(Q_{m}^{(a)})^{2}=Q_{m-1}^{(a)}Q_{m+1}^{(a)}
 +Q_{m}^{(a-1)}Q_{m}^{(a+1)} 
 \quad {\rm for} \quad a \in \{1,2,\dots, s-1\} 
 , \nonumber \\
&&(Q_{m}^{(s)})^{2}=Q_{m-1}^{(s)}Q_{m+1}^{(s)}
 +Q_{m}^{(s-1)}Q_{m}^{(s)},
 \label{Q-sys}
\end{eqnarray}
where $m \in {\mathbb Z}_{\ge 1}$; $Q^{(a)}_{0}=Q^{(0)}_{m}=1$. 
We are considering the case without an external field. 
If an external field exists, (\ref{q-sys-sol}) will be deformed. 

We may assume 
\begin{eqnarray}
\widetilde{T}^{(a)}_{m}(v)=Q^{(a)}_{m} +
\sum_{j=1}^{\frac{N}{2}}
\sum_{k=1}^{2}
\left\{
  \frac{b^{(a)}_{m,j,k}}{(v-\tilde{\beta}^{(a)}_{m,k})^{j}}
 +\frac{\overline{b}^{(a)}_{m,j,k}}{(v+\tilde{\beta}^{(a)}_{m,k})^{j}}
\right\},
\label{expan}
\end{eqnarray}
where the coefficients $b^{(a)}_{m,j,k},
\overline{b}^{(a)}_{m,j,k} \in {\mathbb C}$ 
are given as follows
\begin{eqnarray}
&& b^{(a)}_{m,j,k}= \oint_{C^{(a)}_{m,k}} \frac{{\mathrm d} v}{2\pi i}
 \widetilde{T}^{(a)}_{m}(v)(v-\tilde{\beta}^{(a)}_{m,k})^{j-1},\nonumber \\
&& \overline{b}^{(a)}_{m,j,k}=
 \oint_{\overline{C}^{(a)}_{m,k}} \frac{{\mathrm d} v}{2\pi i}
 \widetilde{T}^{(a)}_{m}(v)(v+\tilde{\beta}^{(a)}_{m,k})^{j-1}.
 \label{coeff}
\end{eqnarray}
Here the contour $C^{(a)}_{m,k}$ must surround $\tilde{\beta}^{(a)}_{m,k}$ 
counterclockwise manner and 
must not contain $\tilde{\beta}^{(a)}_{m,p}$ ($p \ne k$),
$-\tilde{\beta}^{(a)}_{m,1},-\tilde{\beta}^{(a)}_{m,2}$; 
the contour $\overline{C}^{(a)}_{m,k}$ must surround $-\tilde{\beta}^{(a)}_{m,k}$ 
counterclockwise manner and 
must not contain $-\tilde{\beta}^{(a)}_{m,p}$ ($p \ne k$),
$\tilde{\beta}^{(a)}_{m,1},\tilde{\beta}^{(a)}_{m,2}$.
Using the $T$-system (\ref{T-system}), we can modify (\ref{coeff}) as 
\begin{eqnarray}
&& b^{(a)}_{m,j,k}= \oint_{C^{(a)}_{m,k}} \frac{{\mathrm d} v}{2\pi i}
 \bigg\{
 \frac{\widetilde{T}^{(a)}_{m-1}(v-\frac{i}{2})
       \widetilde{T}^{(a)}_{m+1}(v-\frac{i}{2})}
      {\widetilde{T}^{(a)}_{m}(v-i)} \nonumber \\
&& \hspace{80pt} +
 \frac{\widetilde{T}^{(a-1)}_{m}(v-\frac{i}{2})
       \widetilde{T}^{(a+1)}_{m}(v-\frac{i}{2})}
      {\widetilde{T}^{(a)}_{m}(v-i)}
 \bigg\}
 (v-\tilde{\beta}^{(a)}_{m,k})^{j-1},\nonumber \\
&& \overline{b}^{(a)}_{m,j,k}=
 \oint_{\overline{C}^{(a)}_{m,k}} \frac{{\mathrm d} v}{2\pi i}
\bigg\{
 \frac{\widetilde{T}^{(a)}_{m-1}(v+\frac{i}{2})
       \widetilde{T}^{(a)}_{m+1}(v+\frac{i}{2})}
      {\widetilde{T}^{(a)}_{m}(v+i)} \nonumber \\
&& \hspace{80pt} +
 \frac{\widetilde{T}^{(a-1)}_{m}(v+\frac{i}{2})
       \widetilde{T}^{(a+1)}_{m}(v+\frac{i}{2})}
      {\widetilde{T}^{(a)}_{m}(v+i)}
 \bigg\}
 (v+\tilde{\beta}^{(a)}_{m,k})^{j-1}, 
 \label{coeff2}
\end{eqnarray}
where $\widetilde{T}^{(s+1)}_{m}(v):=\widetilde{T}^{(s)}_{m}(v)$. 
The first term and the second term in the 
first (resp. second) bracket $\{\cdots \}$ in (\ref{coeff2}) 
have common poles at 
$\tilde{z}^{(a)}_{m}+i$ (resp. $\tilde{z}^{(a)}_{m}-i$). 
However, these common poles are spurious since 
$\widetilde{T}^{(a)}_{m}(v)$ 
has no pole at these points. 
We also note that a subsidiary condition 
$Y^{(a)}_{m}(\tilde{z}^{(a)}_{m}\pm \frac{i}{2})=-1$ for the excited state 
TBA equation (see, p.2343 in \cite{ST00} and (\ref{Y-fun})) 
follows from this observation. 
Substituting (\ref{coeff2}) into (\ref{expan}), we obtain
\begin{eqnarray}
&& \hspace{-20pt}
\widetilde{T}^{(a)}_{m}(v)=Q^{(a)}_{m} \nonumber \\ 
&& +
\sum_{k=1}^{2}
\oint_{C^{(a)}_{m,k}} \frac{{\mathrm d} y}{2\pi i} 
\frac{1-\left(\frac{y}{v-\tilde{\beta}^{(a)}_{m,k}}\right)^{\frac{N}{2}}}
 {v-y-\tilde{\beta}^{(a)}_{m,k}}
 \bigg\{ 
 \frac{\widetilde{T}^{(a)}_{m-1}(y+\tilde{\beta}^{(a)}_{m,k}-\frac{i}{2}) 
 \widetilde{T}^{(a)}_{m+1}(y+\tilde{\beta}^{(a)}_{m,k}-\frac{i}{2})}
 {\widetilde{T}^{(a)}_{m}(y+\tilde{\beta}^{(a)}_{m,k}-i)} \nonumber \\
 && \hspace{90pt} +
 \frac{\widetilde{T}^{(a-1)}_{m}(y+\tilde{\beta}^{(a)}_{m,k}-\frac{i}{2}) 
 \widetilde{T}^{(a+1)}_{m}(y+\tilde{\beta}^{(a)}_{m,k}-\frac{i}{2})}
 {\widetilde{T}^{(a)}_{m}(y+\tilde{\beta}^{(a)}_{m,k}-i)}
 \bigg\} \nonumber \\
&& +
\sum_{k=1}^{2}
\oint_{\overline{C}^{(a)}_{m,k}} \frac{{\mathrm d} y}{2\pi i} 
 \frac{1-\left(\frac{y}{v+\tilde{\beta}^{(a)}_{m,k}}\right)^{\frac{N}{2}}}
 {v-y+\tilde{\beta}^{(a)}_{m,k}}
 \bigg\{ 
 \frac{\widetilde{T}^{(a)}_{m-1}(y-\tilde{\beta}^{(a)}_{m,k}+\frac{i}{2}) 
 \widetilde{T}^{(a)}_{m+1}(y-\tilde{\beta}^{(a)}_{m,k}+\frac{i}{2})}
 {\widetilde{T}^{(a)}_{m}(y-\tilde{\beta}^{(a)}_{m,k}+i)} \nonumber \\
 && \hspace{90pt} +
 \frac{\widetilde{T}^{(a-1)}_{m}(y-\tilde{\beta}^{(a)}_{m,k}+\frac{i}{2}) 
 \widetilde{T}^{(a+1)}_{m}(y-\tilde{\beta}^{(a)}_{m,k}+\frac{i}{2})}
 {\widetilde{T}^{(a)}_{m}(y-\tilde{\beta}^{(a)}_{m,k}+i)}
 \bigg\}, \nonumber \\ 
 && \hspace{100pt} 
 {\rm for} \quad a \in \{1,2,\dots,s\}
 \quad {\rm and} \quad m \in {\mathbb Z}_{\ge 1},
 \label{nlie1}
\end{eqnarray}
where $\widetilde{T}^{(s+1)}_{m}(v)=\widetilde{T}^{(s)}_{m}(v)$.
Here the contour $C^{(a)}_{m,k}$ (resp. $\overline{C}^{(a)}_{m,k}$) 
must surround $0$ 
counterclockwise manner and 
must not contain 
$\tilde{\beta}^{(a)}_{m,p}-\tilde{\beta}^{(a)}_{m,k}$ ($p \ne k$),
$-\tilde{\beta}^{(a)}_{m,1}-\tilde{\beta}^{(a)}_{m,k},
-\tilde{\beta}^{(a)}_{m,2}-\tilde{\beta}^{(a)}_{m,k}$ 
(resp. $-\tilde{\beta}^{(a)}_{m,p}+\tilde{\beta}^{(a)}_{m,k}$ ($p \ne k$),
$\tilde{\beta}^{(a)}_{m,1}+\tilde{\beta}^{(a)}_{m,k},
\tilde{\beta}^{(a)}_{m,2}+\tilde{\beta}^{(a)}_{m,k}$).
%
For $m \ge 2$, 
the first term and the second term in the first bracket $\{ \cdots \}$ in 
(\ref{nlie1}) have a common singularity at $0$ which contributes to 
the contour integral. 
On the other hand for $m=1$, this 
singularity at $0$ from the first term disappears since 
$\widetilde{T}^{(a)}_{0}(y)=1$. Thus 
the contribution to the contour integral from 
the first term in the first bracket $\{ \cdots \}$ in 
(\ref{nlie1}) vanishes as long as 
 the contour $C^{(a)}_{1,k}$ does not contain the 
singularities at $\tilde{z}^{(a)}_{1}-\tilde{\beta}^{(a)}_{1,k}+i$ 
(cf. Conjecture \ref{conj}). 
This situation is parallel with the second bracket $\{ \cdots \}$ in 
(\ref{nlie1}).
Therefore for $m=1$, (\ref{nlie1}) reduces to 
\begin{eqnarray}
&& \hspace{-20pt}
\widetilde{T}^{(a)}_{1}(v)=Q^{(a)}_{1} \nonumber \\ 
&& +
\sum_{k=1}^{2}
\oint_{C^{(a)}_{1,k}} \frac{{\mathrm d} y}{2\pi i} 
\frac{1-\left(\frac{y}{v-\tilde{\beta}^{(a)}_{1,k}}\right)^{\frac{N}{2}}}
 {v-y-\tilde{\beta}^{(a)}_{1,k}}
 \frac{\widetilde{T}^{(a-1)}_{1}(y+\tilde{\beta}^{(a)}_{1,k}-\frac{i}{2}) 
 \widetilde{T}^{(a+1)}_{1}(y+\tilde{\beta}^{(a)}_{1,k}-\frac{i}{2})}
 {\widetilde{T}^{(a)}_{1}(y+\tilde{\beta}^{(a)}_{1,k}-i)}
 \nonumber \\
&& +
\sum_{k=1}^{2}
\oint_{\overline{C}^{(a)}_{1,k}} \frac{{\mathrm d} y}{2\pi i} 
 \frac{1-\left(\frac{y}{v+\tilde{\beta}^{(a)}_{1,k}}\right)^{\frac{N}{2}}}
 {v-y+\tilde{\beta}^{(a)}_{1,k}}
 \frac{\widetilde{T}^{(a-1)}_{1}(y-\tilde{\beta}^{(a)}_{1,k}+\frac{i}{2}) 
 \widetilde{T}^{(a+1)}_{1}(y-\tilde{\beta}^{(a)}_{1,k}+\frac{i}{2})}
 {\widetilde{T}^{(a)}_{1}(y-\tilde{\beta}^{(a)}_{1,k}+i)}
 \nonumber \\ 
 && \hspace{180pt} 
 {\rm for} \quad a \in \{1,2,\dots,s\}, \label{nlie2} 
\end{eqnarray}
where $\widetilde{T}^{(s+1)}_{1}(v)=\widetilde{T}^{(s)}_{1}(v)$. 
Here the contour $C^{(a)}_{1,k}$ (resp. $\overline{C}^{(a)}_{1,k}$) 
must surround $0$ 
counterclockwise manner and 
must not contain 
$\tilde{z}^{(a)}_{1}-\tilde{\beta}^{(a)}_{1,k}+i$, 
$\tilde{\beta}^{(a)}_{1,p}-\tilde{\beta}^{(a)}_{1,k}$ ($p \ne k$),
$-\tilde{\beta}^{(a)}_{1,1}-\tilde{\beta}^{(a)}_{1,k},
-\tilde{\beta}^{(a)}_{1,2}-\tilde{\beta}^{(a)}_{1,k}$ 
(resp. 
$\tilde{z}^{(a)}_{1}+\tilde{\beta}^{(a)}_{1,k}-i$, 
$-\tilde{\beta}^{(a)}_{1,p}+\tilde{\beta}^{(a)}_{1,k}$ ($p \ne k$),
$\tilde{\beta}^{(a)}_{1,1}+\tilde{\beta}^{(a)}_{1,k},
\tilde{\beta}^{(a)}_{1,2}+\tilde{\beta}^{(a)}_{1,k}$).
Now we shall take the Trotter limit $N \to \infty$ in (\ref{nlie1}). 
\begin{eqnarray}
&& \hspace{-20pt}
\mathcal{T}^{(a)}_{m}(v)=Q^{(a)}_{m} \nonumber \\ 
&& +
\sum_{k=1}^{2}
\oint_{C^{(a)}_{m,k}} \frac{{\mathrm d} y}{2\pi i} 
\frac{1}{v-y-\beta^{(a)}_{m,k}}
 \bigg\{ 
 \frac{\mathcal{T}^{(a)}_{m-1}(y+\beta^{(a)}_{m,k}-\frac{i}{2}) 
 \mathcal{T}^{(a)}_{m+1}(y+\beta^{(a)}_{m,k}-\frac{i}{2})}
 {\mathcal{T}^{(a)}_{m}(y+\beta^{(a)}_{m,k}-i)} \nonumber \\
 && \hspace{90pt} +
 \frac{\mathcal{T}^{(a-1)}_{m}(y+\beta^{(a)}_{m,k}-\frac{i}{2}) 
 \mathcal{T}^{(a+1)}_{m}(y+\beta^{(a)}_{m,k}-\frac{i}{2})}
 {\mathcal{T}^{(a)}_{m}(y+\beta^{(a)}_{m,k}-i)}
 \bigg\} \nonumber \\
&& +
\sum_{k=1}^{2}
\oint_{\overline{C}^{(a)}_{m,k}} \frac{{\mathrm d} y}{2\pi i} 
 \frac{1}
 {v-y+\beta^{(a)}_{m,k}}
 \bigg\{ 
 \frac{\mathcal{T}^{(a)}_{m-1}(y-\beta^{(a)}_{m,k}+\frac{i}{2}) 
 \mathcal{T}^{(a)}_{m+1}(y-\beta^{(a)}_{m,k}+\frac{i}{2})}
 {\mathcal{T}^{(a)}_{m}(y-\beta^{(a)}_{m,k}+i)} \nonumber \\
 && \hspace{90pt} +
 \frac{\mathcal{T}^{(a-1)}_{m}(y-\beta^{(a)}_{m,k}+\frac{i}{2}) 
 \mathcal{T}^{(a+1)}_{m}(y-\beta^{(a)}_{m,k}+\frac{i}{2})}
 {\mathcal{T}^{(a)}_{m}(y-\beta^{(a)}_{m,k}+i)}
 \bigg\}
 \nonumber \\ 
 && \hspace{100pt} 
 {\rm for} \quad a \in \{1,2,\dots,s\}
 \quad {\rm and} \quad m \in {\mathbb Z}_{\ge 1},
 \label{nlie3}
\end{eqnarray}
where $\mathcal{T}^{(a)}_{m}(v):=
\lim_{N \to \infty} \widetilde{T}^{(a)}_{m}(v)$; 
$\beta^{(a)}_{m,k}:=\lim_{N \to \infty} \tilde{\beta}^{(a)}_{m,k}$ 
($\beta^{(a)}_{m,1}=\frac{m+a}{2}i$, 
$\beta^{(a)}_{m,2}=\frac{g+m-a}{2}i$); 
$\mathcal{T}^{(s+1)}_{m}(v)=\mathcal{T}^{(s)}_{m}(v)$; 
$\mathcal{T}^{(a)}_{0}(v)=1$; $\mathcal{T}^{(0)}_{m}(v)$ 
is a known function:
\begin{eqnarray}
\mathcal{T}^{(0)}_{m}(v)=
\lim_{N \to \infty} \widetilde{T}^{(0)}_{m}(v)= 
        \exp \left(-\frac{mJ}{(v^2+\frac{m^2}{4})T}\right).
\end{eqnarray}
Here the contour $C^{(a)}_{m,k}$ (resp. $\overline{C}^{(a)}_{m,k}$) 
must surround $0$ 
counterclockwise manner with the condition 
$|y| < |v-\beta^{(a)}_{m,k}|$ 
(resp. $|y| < |v+\beta^{(a)}_{m,k}|$)
and 
must not contain 
$\beta^{(a)}_{m,p}-\beta^{(a)}_{m,k}$ ($p \ne k$),
$-\beta^{(a)}_{m,1}-\beta^{(a)}_{m,k},
-\beta^{(a)}_{m,2}-\beta^{(a)}_{m,k}$ 
(resp. $-\beta^{(a)}_{m,p}+\beta^{(a)}_{m,k}$ ($p \ne k$),
$\beta^{(a)}_{m,1}+\beta^{(a)}_{m,k},
\beta^{(a)}_{m,2}+\beta^{(a)}_{m,k}$).
In particular for $m=1$, we have
\begin{eqnarray}
{\mathcal T}^{(a)}_{1}(v)=Q^{(a)}_{1} 
&+&
\sum_{k=1}^{2}
\oint_{C^{(a)}_{1,k}} \frac{{\mathrm d} y}{2\pi i} 
 \frac{\mathcal{T}^{(a-1)}_{1}(y+\beta^{(a)}_{1,k}-\frac{i}{2}) 
 \mathcal{T}^{(a+1)}_{1}(y+\beta^{(a)}_{1,k}-\frac{i}{2})}
 {(v-y-\beta^{(a)}_{1,k})\mathcal{T}^{(a)}_{1}(y+\beta^{(a)}_{1,k}-i)}
 \nonumber \\
&+&
\sum_{k=1}^{2}
\oint_{\overline{C}^{(a)}_{1,k}} \frac{{\mathrm d} y}{2\pi i} 
 \frac{\mathcal{T}^{(a-1)}_{1}(y-\beta^{(a)}_{1,k}+\frac{i}{2}) 
 \mathcal{T}^{(a+1)}_{1}(y-\beta^{(a)}_{1,k}+\frac{i}{2})}
 {(v-y+\beta^{(a)}_{1,k})\mathcal{T}^{(a)}_{1}(y-\beta^{(a)}_{1,k}+i)}
 \nonumber \\ 
 && \hspace{120pt} 
 {\rm for} \quad a \in \{1,2,\dots,s\},
 \label{nlie4}
\end{eqnarray}
where $\mathcal{T}^{(s+1)}_{1}(v)=\mathcal{T}^{(s)}_{1}(v)$. 
Here the contour $C^{(a)}_{1,k}$ (resp. $\overline{C}^{(a)}_{1,k}$) 
must surround $0$ 
counterclockwise manner with the condition 
$|y| < |v-\beta^{(a)}_{1,k}|$ 
(resp. $|y| < |v+\beta^{(a)}_{1,k}|$) and 
must not contain 
$z^{(a)}_{1}-\beta^{(a)}_{1,k}+i$, 
$\beta^{(a)}_{1,p}-\beta^{(a)}_{1,k}$ ($p \ne k$),
$-\beta^{(a)}_{1,1}-\beta^{(a)}_{1,k},
-\beta^{(a)}_{1,2}-\beta^{(a)}_{1,k}$ 
(resp. 
$z^{(a)}_{1}+\beta^{(a)}_{1,k}-i$, 
$-\beta^{(a)}_{1,p}+\beta^{(a)}_{1,k}$ ($p \ne k$),
$\beta^{(a)}_{1,1}+\beta^{(a)}_{1,k},
\beta^{(a)}_{1,2}+\beta^{(a)}_{1,k}$): 
$z^{(a)}_{1}=\lim_{N\to \infty }{\tilde z}^{(a)}_{1}$.
In (\ref{nlie4}), (\ref{q-sys-sol}) is written as a binomial 
coefficient: $Q^{(a)}_{1}=\binom{2s+1}{a}$. 
We note the fact that this set of NLIE (\ref{nlie4}) 
contains only a {\em finite} number of unknown 
functions $\{{\mathcal T}^{(a)}_{1}(v) \}_{1\le a \le s}$.
Moreover, we can express 
$\{{\mathcal T}^{(a)}_{m}(v) \}_{1\le a \le s;m\in{\mathbb Z}_{\ge 1}}$ 
 in terms of the solution of (\ref{nlie4}) 
$\{{\mathcal T}^{(a)}_{1}(v) \}_{1\le a \le s}$ 
by using a Jacobi-Trudi determinant formula \cite{T99}. 
In this sense, we need not consider (\ref{nlie3}) 
for $m \in {\mathbb Z}_{\ge 2}$ 
 in practical calculations. 
 However, consideration on (\ref{nlie3}) for $m \in {\mathbb Z}_{\ge 2}$ 
 manifests the relation between our new NLIE and the 
 traditional TBA equations (see, section 5). 
The free energy per site is given as 
\begin{eqnarray}
f=-J-T\log \mathcal{T}^{(1)}_{1}(0). 
\label{free-en}
\end{eqnarray}
Solving (\ref{nlie4}) iteratively, we can obtain the free 
energy through (\ref{free-en}). 
For $s=1$ case, we find (\ref{nlie4}) converges numerically 
at least for $|T/J|>0.37$.
\section{High Temperature Expansion: $osp(1|2)$ case}
\begin{figure}
\begin{center}
\includegraphics
{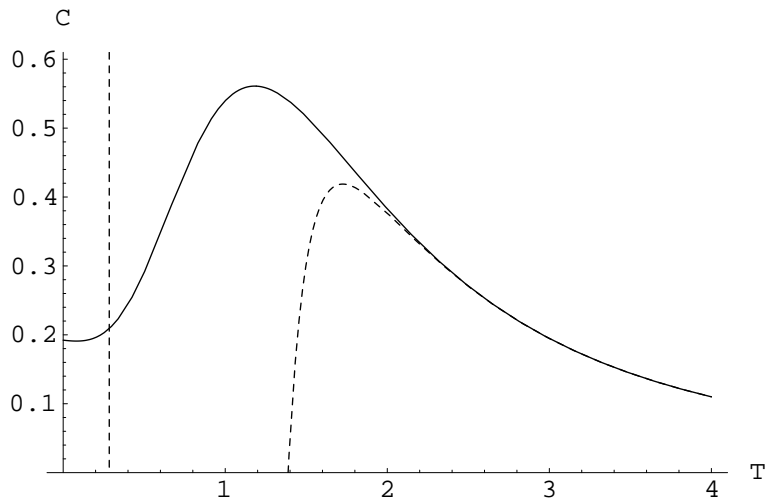}
\end{center}
\caption{Temperature dependence of the high temperature 
expansion of the specific heat for $s=1$ and $J=-1$: 
 the broken line denotes the plain series (\ref{specific-ex}); 
the smooth line denotes its pade approximation.}
\label{specific}
\end{figure}

It was pointed out \cite{ShT02} 
that Takahashi's NLIE 
for the $XXX$-model \cite{Ta01} is very useful to calculate 
the high temperature expansion of the free energy. 
In this section, we shall calculate the high temperature expansion of 
the free energy (\ref{free-en}) for $s=1$ 
by our new NLIE (\ref{nlie4}).  
We assume the following expansion for large $T$: 
\begin{eqnarray}
\mathcal{T}^{(1)}_{1}(v)=
 \exp \left(\sum_{n=0}^{\infty}a_{n}(v)(\frac{J}{T})^{n} \right).
 \label{t-expan}
\end{eqnarray}
Due to the shift of the argument $y$ of 
${\mathcal T}_{1}^{(1)}(y)$  
in (\ref{nlie4}), 
we have to take into account the residues from the 
coefficients $\{a_{n}(v)\}$, which contrasts with the $XXX$-model 
case \cite{ShT02}. 
Thus the derivation is not so easy as the $XXX$-model case. 
But we still expect that 
it is easier than to use the traditional TBA equation which 
contains an infinite number of unknown functions
(as for the high temperature expansion for the $XXX$-model from 
 the traditional TBA equation, 
see for example, p.124 in  \cite{Ta99}). 
To calculate the coefficients $\{a_{n}(v)\}$ efficiently, 
we need further assumption for $n \in {\mathbb Z}_{\ge 1}$: 
\begin{eqnarray}
a_{n}(v)=\sum_{j=0}^{n-1}
\left\{ 
 \frac{b_{n,j}v^{2j}}{(v^2+1)^n}+
 \frac{c_{n,j}v^{2j}}{(v^2+\frac{9}{4})^n}
\right\},
\end{eqnarray}
where $b_{n,j},c_{n,j} \in {\mathbb C}$ are independent of $v$.
Substituting (\ref{t-expan}) into (\ref{nlie4}), we 
obtain the coefficients $\{a_{n}(v)\}$ up to order $12$. 
For example, we have
\begin{eqnarray}
&& a_{0}(v)=\log 3, \nonumber \\
&& a_{1}(v)=-\frac{2}{3(v^2+1)}-\frac{1}{3(v^2+\frac{9}{4})},
 \nonumber \\
&& a_{2}(v)=\frac{4 \left( 77 + 32\,v^2 \right) }{405(v^2+1)^2}-
 \frac{4 \left( 27 + 32\,v^2 \right) }{405(v^2+\frac{9}{4})^2},\nonumber \\
&& a_{3}(v)=-\frac{8 \left( 727 + 1040\,v^2 + 448\,v^4 \right) }
            {10935(v^2+1)^3}+
\frac{28 \left( 837 + 720\,v^2 + 128\,v^4 \right) }
     {10935(v^2+\frac{9}{4})^3},\nonumber \\ 
&& a_{4}(v)=
\frac{2 \left( -9097 + 781314\,v^2 + 783969\,v^4 + 175808\,v^6 \right)
      }{7381125(v^2+1)^4} \nonumber \\ 
&& \hspace{30pt} -
\frac{144376263 + 138298104\,v^2 + 37692144\,v^4 + 2812928\,v^6}
  {59049000(v^2+\frac{9}{4})^4}.
\end{eqnarray}
Using $\{a_{n}(0)\}$, we obtain 
\begin{eqnarray}
&& \hspace{-20pt} \frac{f}{T}=
-\log 3 - \frac{5\,J}{27\,T} - 
\frac{172\,J^2}{243\,T^2} + 
\frac{20296\,J^3}{59049\,T^3} + 
\frac{52010\,J^4}{531441\,T^4} - 
\frac{466964\,J^5}{1594323\,T^5} 
\nonumber \\ &&+ 
\frac{252697291\,J^6}{1937102445\,T^6} + 
\frac{2867981638\,J^7}{17433922005\,T^7} - 
\frac{319036008559\,J^8}{1255242384360\,T^8} 
\nonumber \\ && + 
\frac{68608132529023\,J^9}{1921650566216724\,T^9} + 
\frac{69872025931694239\,J^{10}}{288247584932508600\,T^{10}} 
 \\ && - 
\frac{157451559799196839\,J^{11}}{648557066098144350\,T^{11}} - 
\frac{48290843858722808551\,J^{12}}{693437215072135939020\,T^{12}} + 
 O((\frac{J}{T})^{13}). \nonumber
\end{eqnarray}
We also calculate the 
specific heat $C=-T\frac{\partial^{2} f}{\partial T^{2}} $.
\begin{eqnarray}
&& \hspace{-20pt} C=
\frac{344\,J^2}{243\,T^2} - 
\frac{40592\,J^3}{19683\,T^3} - 
\frac{208040\,J^4}{177147\,T^4} + 
\frac{9339280\,J^5}{1594323\,T^5} - 
\frac{505394582\,J^6}{129140163\,T^6} \nonumber \\ 
&& - 
\frac{40151742932\,J^7}{5811307335\,T^7} + 
\frac{2233252059913\,J^8}{156905298045\,T^8} - 
\frac{137216265058046\,J^9}{53379182394909\,T^9} \nonumber \\ 
&&- 
\frac{69872025931694239\,J^{10}}{3202750943694540\,T^{10}} + 
\frac{1731967157791165229\,J^{11}}{64855706609814435\,T^{11}}  
\nonumber \\  &&+ 
\frac{48290843858722808551\,J^{12}}{5253312235394969235\,T^{12}} + 
 O((\frac{J}{T})^{13}). \label{specific-ex}
\end{eqnarray}
We have plotted the high temperature expansion of the 
specific heat (\ref{specific-ex}) 
in figure \ref{specific}. 
For comparison, 
we also plotted a pade approximation for (\ref{specific-ex}). 
For large $T$, (\ref{specific-ex}) agrees with our TBA analysis 
(see, Fig.1 in  \cite{ST01}). 
This indicates the validity of our new NLIE (\ref{nlie4}).
\section{Discussion}
In this letter, we have derived a system of NLIE 
with a {\em finite} number of unknown functions, 
which describes thermodynamics of the $osp(1|2s)$ integrable
 spin chan. This type of NLIE 
 for {\em arbitrary} rank is derived for the first time. 
We shall point out a relation between 
our new NLIE and 
usual TBA equations. 
The functions $\{ \widetilde{T}^{(a)}_{m}(v)\}$ 
are related to the dependent variables 
$\{ Y^{(a)}_{m}(v)\}$
 of the $Y$-system (or TBA equations) \cite{T02} as 
\begin{eqnarray}
&& Y^{(a)}_{m}(v)=
 \frac{\widetilde{T}^{(a)}_{m+1}(v)
 \widetilde{T}^{(a)}_{m-1}(v)}
 {\widetilde{T}^{(a-1)}_{m}(v)\widetilde{T}^{(a+1)}_{m}(v)},
 \quad 
 Y^{(s)}_{m}(v)=
 \frac{\widetilde{T}^{(s)}_{m+1}(v)
 \widetilde{T}^{(s)}_{m-1}(v)}
 {\widetilde{T}^{(s-1)}_{m}(v)
 \widetilde{T}^{(s)}_{m}(v)}
 \label{Y-fun} \\
&& \hspace{70pt} {\rm for} \quad a \in \{1,2,\dots,s-1\}
 \quad {\rm and} \quad m \in {\mathbb Z}_{\ge 1}.
\nonumber 
\end{eqnarray}
This relation is also valid in the Trotter limit. 

In closing this letter, we shall enumerate future problems:\\
$\bullet$ There are $T$-systems for other algebras 
\cite{KNS95,KS95-2,T97,T98,T98-2,T99-1,T99}. 
Using these $T$-systems, we can 
 derive NLIE 
similar to the ones in this letter,
 which will be reported elsewhere \cite{T02-2}. \\
$\bullet$ In this letter, we have considered the case 
without an external field. When an external field 
exists, a fugacity expansion 
of the free energy can be calculated recursively 
through (\ref{nlie4}) 
(as for the fugacity expansion of the $XXX$-model from 
 the traditional TBA equation, 
see for example, p.127 in  \cite{Ta99}).\\
$\bullet$
One will be able to extend (\ref{nlie4}) to 
trigonometric or elliptic case.  
In this case, one must take into account 
the periodicity in 
 the summation in (\ref{expan}). 
\section*{Acknowledgments}
\noindent
The author is financially supported by 
Inoue Foundation for Science. 
  
\end{document}